\documentclass[12pt]{article}
\usepackage{a4wide, amsmath, latexsym, epsfig,amsfonts,
  psfrag,graphicx, rotating, fancyhdr, tabularx, afterpage,booktabs}
\usepackage{amssymb}
\usepackage[bookmarks=true,colorlinks=true,citecolor=black,linkcolor=black,urlcolor=black]{hyperref}

\newcommand\ba{\begin{eqnarray}}
\newcommand\ea{\end{eqnarray}}

\newcommand{\GeV}{~\mbox{GeV}}

\def\phist{\ensuremath{\phi^{*}_{\eta}}}
\def\zpt{\ensuremath{p^{Z}_{T}}}
\def\diffx{\ensuremath{1/\sigma^{{\rm fid}} \cdot d\sigma^{{\rm fid}}/d\phi^*_\eta}}
\begin{document}
\begin{titlepage}
 
\begin{flushright} 
{ IFJPAN-IV-2013-2 \\ CERN-PH-TH/2013-040} 
\end{flushright}

\vspace{0.2cm} 
\begin{center}
{\Huge \bf  Observable $\phist$  at LHC and second-order QED matrix element in 
 $Z/\gamma^*\rightarrow l^+l^-$ decays}
\end{center}

\vspace{4mm}

\begin{center}
   {\bf    Thi Kieu Oanh Doan$^{a}$,    W.P{\l}aczek$^b$  and Z. W\c{a}s$^{c,d}$}\\
\vspace{3mm}
{\em $^a$LAPP Annecy, CNRS, France.}\\
\vspace{1mm}
{\em $^b$Marian Smoluchowski Institute of Physics, Jagiellonian University,\\
         ul.\ Reymonta 4, 30-059 Krakow, Poland.} \\
\vspace{1mm}
{\em $^c$Institute of Nuclear Physics, PAN,
         ul. Radzikowskiego 152, Krakow, Poland.}\\
\vspace{1mm}
{\em $^d$CERN PH-TH, CH-1211 Geneva 23, Switzerland.}

\end{center}

\vspace{4mm}
\begin{abstract}
In a recent publication by  ATLAS collaboration a new 
observable, the so-called $\phist$ angle, was  used for precise
measurement of transverse $Z$ momentum.
One of the dominant systematic errors for this measurement originates from the  
theoretical control of QED final-state bremsstrahlung.
At present, it is estimated at the $0.3\%$ level for the shape of the $\phist$ distribution. In this paper
we discuss the possible effects of the second-order QED matrix element
for that quantity. For that purpose, results from simulations based on the
Yennie--Frautchi--Suura (YFS) exponentiation and featuring the second-order matrix 
elements are  used and compared with the case when the matrix element 
is restricted to the first order. 
From this study we conclude that in order to reach the precision below $0.3\%$ 
for the  $\phist$ distribution at the LHC, inclusion of the second-order QED matrix element 
in a respective Monte Carlo event generator is necessary. 

\end{abstract}

 
 \vspace{8mm}
\begin{flushleft}
{IFJPAN-IV-2013-2 \\  CERN-PH-TH/2013-040\\
 March, 2013}
\end{flushleft}

\end{titlepage}

\section {Introduction}
The main purpose of LHC experiments is to search for new elementary particles and interactions. The Higgs-like particle observability papers \cite{:2012gk,:2012gu} document a fundamental achievement for  this goal.
Another important class of the  LHC measurements aims on high-precision consistency tests of the
Standard Model. In this respect, a precise measurement of the $W$ mass
plays a particularly important role \cite{Besson:2009zzb}. To reduce 
systematic errors, 
improvements in the measurement techniques are desirable.
It is generally believed that 
higher precision can be  achieved thanks to the use, whenever possible, of leptonic
degrees of freedom rather than hadronic ones.  In this way  the precision 
better by even an order of magnitude can be achieved. 
At present, the best measurements of $W$ mass 
have been performed in $p \bar p$ collisions at Tevatron~\cite{Aaltonen:2009aa,Aaltonen:2012bp,PhysRevLett.108.151804}.
In the CDF collaboration the precision of $19\,$MeV on the $W$ mass was achieved. They have estimated that
the largest contribution from theory to the systematic error originates  from
 parton distribution functions ({\tt PDF}'s) and initial-state hadronic interactions in general. 

An approach~\cite{Vesterinen:2009,Banfi:2011,Banfi:2011dx} based on
measuring the  $\phist$ angle, instead of  $Z$ transverse momentum ($\zpt$), may offer a significant improvement. 
In Ref.~\cite{phist} the first results of $\phist$ measurements with ATLAS 
experiment were presented already. In that experimental publication
the systematic error of $0.3\%$ to $\phist$ due to implementation 
of QED final state radiation (FSR) in the Monte Carlo generators  
 used, was estimated in proportion of differences  observed 
between {\tt PHOTOS} 
\cite{Golonka:2005pn,Golonka:2006tw,Barberio:1994qi} and {\tt SHERPA} \cite{Gleisberg:2008ta} predictions. 
If one  understands the pattern of these 
differences, one can hopefully reduce systematic errors further. In this sense
presented here work 
is continuation of recent efforts in this direction documented in Ref.~\cite{Arbuzov:2012dx}.

One of the possible sources of the above differences may be  the second-order matrix element 
for QED final-state radiation (FSR) in $Z$ decays.  Such a matrix element is missing in {\tt PHOTOS} and 
in {\tt SHERPA}, but its dominant contribution is taken into account
thanks to iterative and/or multiphoton nature of the algorithms. Multitude of tests for {\tt PHOTOS}
were devoted to this point over time \cite{Golonka:2006tw,Arbuzov:2012dx,erw:1994}. 
In particular, it was shown \cite{Golonka:2005pn} that despite the fact that {\tt PHOTOS}
is not using  a matrix-element-based kernel, it agrees substantially better 
with distributions obtained from the {\tt KKMC} Monte Carlo~\cite{kkcpc:1999} 
if exclusive exponentiation featuring 
the second-order matrix element is used rather than if the matrix element is restricted to the  first-order
only. The {\tt KKMC} program is particularly suitable as a source of precise
numerical benchmarks, because its precision was studied at the LEP time in a great 
detail, see e.g. Ref.~\cite{Jadach:2000ir}, and with precision requirements surpassing the
present-day requirements of the LHC experiments.  
It is known since more than 20 years now \cite{yfs3-pl:1992} that  
second-order terms
are necessary for an exponentiation-based Monte Carlo generator if it is expected to assure a high precision in the case of precision observables of $e^+e^-$ colliders. 
That is why such terms were  implemented in the $e^+e^-\to l^+l^-$ generators 
{\tt KORALZ} \cite{koralz4:1994} and  in the {\tt KKMC} program as well.
The  precision of $0.06\%$ was evaluated in \cite{Ward:1998ht} for small-angle Bhabha 
scattering simulated with the  Monte Carlo generator {\tt BHLUMI} \cite{Jadach:1996is}.

For the LHC applications, the second-order QED FSR matrix element 
was found to be important for simulations 
  of background for Higgs-boson searches in the $2\gamma$ channel \cite{erw:1994}. With time and improving experimental precision, contribution from the second-order QED FSR matrix element
 may become important for more inclusive
observables at the LHC as well.
First attempts to 
implement the second-order matrix element (ME) to the LHC Monte Carlo generators are on-going; 
a good example is \cite{Yost:2012az}.
 
Before such projects are completed,
let us provide some quantitative results using the LEP-era Monte Carlo program
{\tt KKMC}. 
For emulation of the initial-state hadronic interactions, parton distribution functions ({\tt PDF}'s) 
combined with the transverse momentum smearing based on 
the {\tt RESBOS} program \cite{Balazs:1997xd} are used only. 
Such an approach doesn't take into account the 
detector simulations, nonetheless it should be sufficient to quantify the size of the effect 
of the second-order ME. 

Our note is organized as follows.  In Section~2 we describe our Monte Carlo set-up. Section 3 is devoted to a careful definition of 
the $\phist$ distribution, including selections criteria, etc. In Section~4 we present our numerical results.
Section 5 summarizes the paper with conclusions.

\section{Monte Carlo set-up} \label{noSM}

The {\tt KKMC} event generator is constructed  for $e^+e^- \to \mu^+ \mu^- (\tau^+\tau^-,\ q\bar q)$ processes and for centre-of-mass system energies from 
the production  threshold up to $200\,$GeV. For our purposes
it was adapted to work for quark-anti-quark annihilation process $q \bar q \to l^+l^-$. However, in this case, only QED FSR can 
be generated. By default, the second-order matrix element is 
active, but the program can be downgraded to the first-order FSR as well. 
We use this program as a building block in our 
simulation chain.
With the help of  {\tt WINHAC}, the MC event generator for Drell--Yan processes \cite{Placzek:2003zg,Placzek:2009jy},
using the {\tt MSTW2008NLO}  \cite{Martin:2009iq} parametrisation of {\tt PDF}'s 
we generate a series of the Bjorken variables $x_1$ and $x_2$ for $pp$ collisions  at $\sqrt{s}=7\,$TeV 
corresponding to production of $Z$-bosons in a very narrow invariant mass range: $M_Z\pm 1\,$MeV.  
Then we generate monochromatic $Z$ decays including QED FSR bremsstrahlung.
Let us stress that we aim at the estimation of a small second-order QED FSR effect and not at the complete predictions. 
That is why we can limit ourselves to the dominant QCD initial-state effects only. 

\begin{figure}[htp!]
\begin{tabular}{ccc}
  \includegraphics[width=0.48\columnwidth]{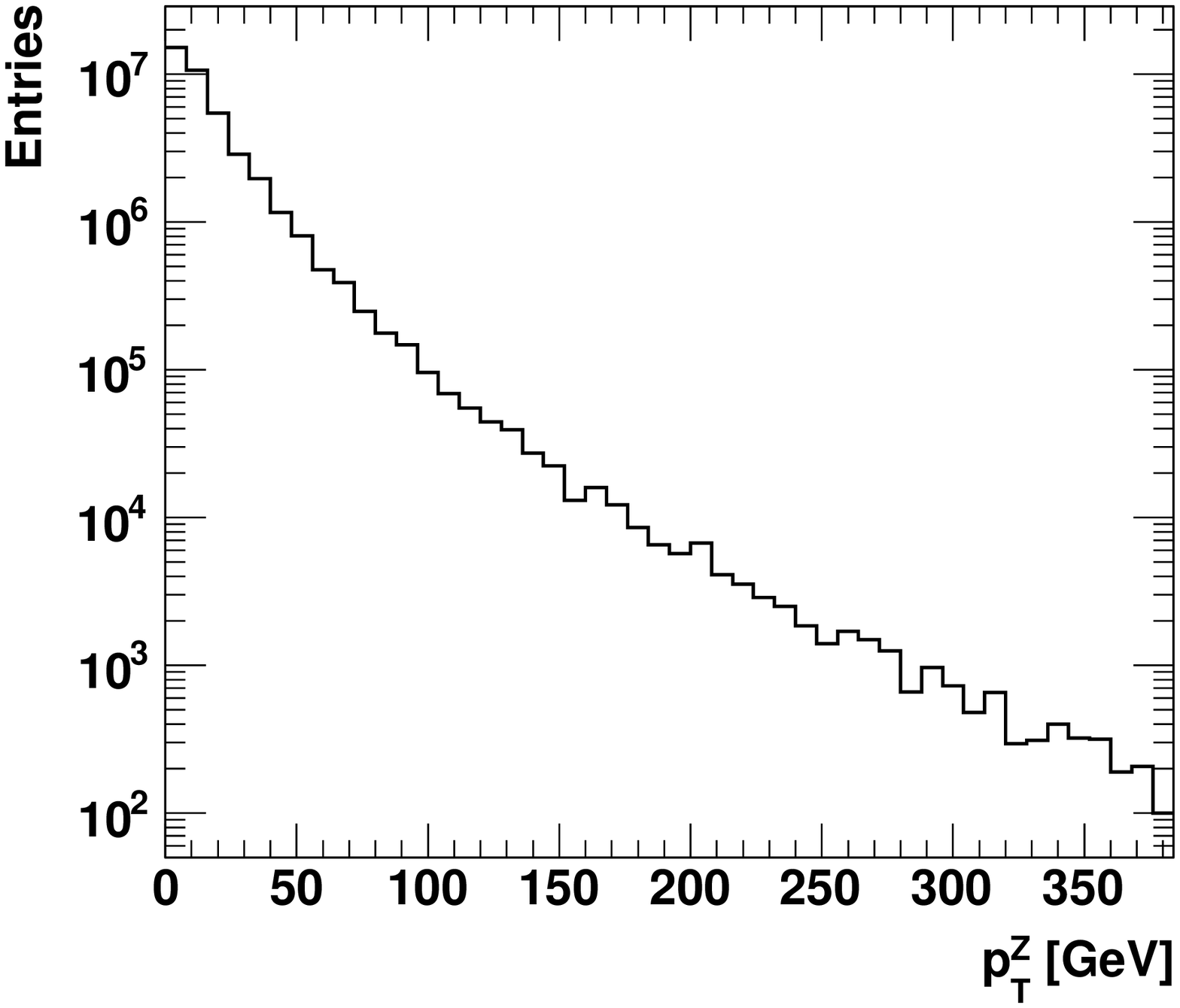} & 
  \includegraphics[width=0.48\columnwidth]{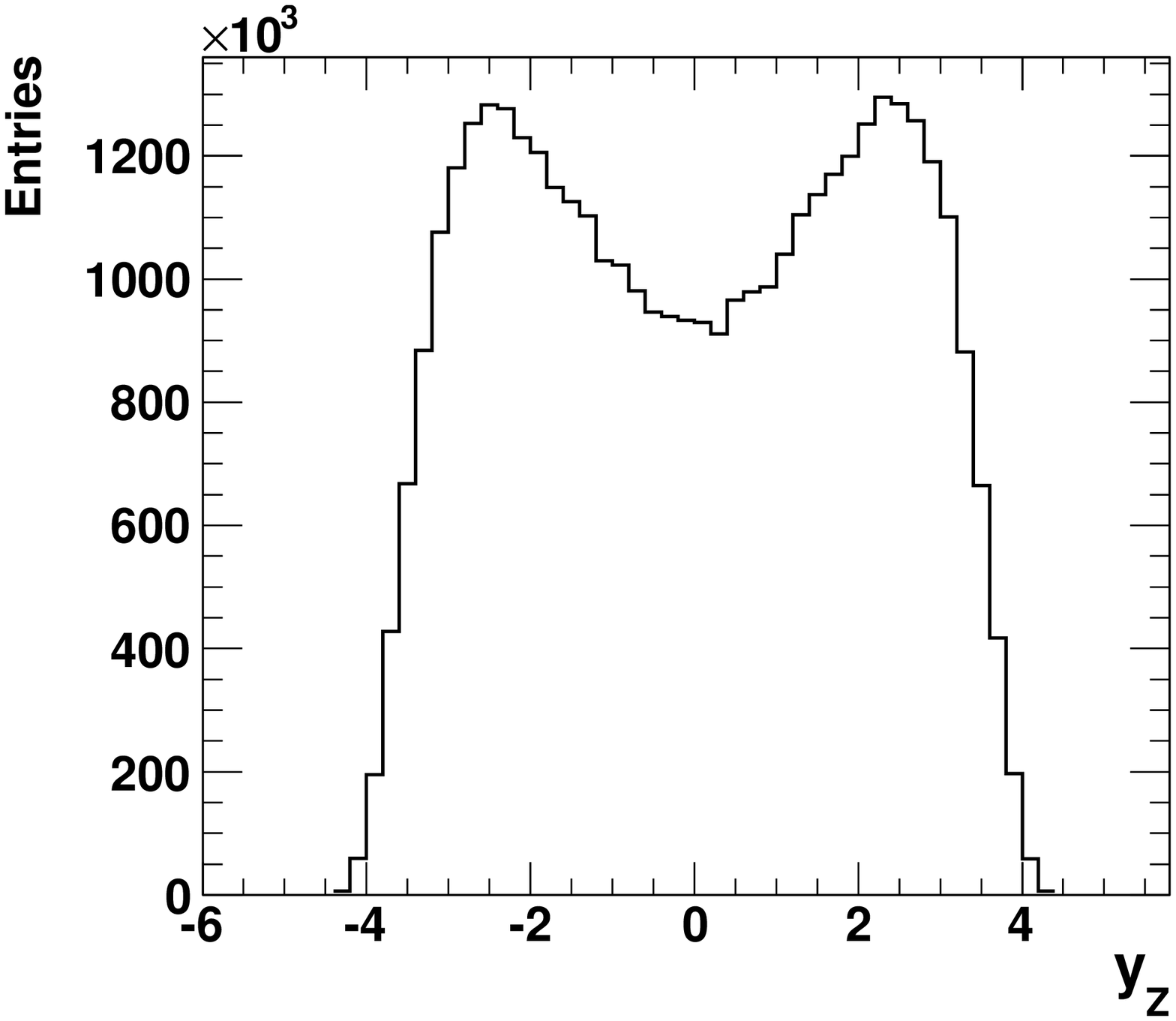} 
  \\ &
\end{tabular}
\caption{%
The distributions of  the $Z$ transverse momentum $p_T^Z$  (left) and rapidity $y_Z$ (right) 
constructed from the Monte Carlo sample used in our study.  No selection
 cuts were applied.
The quark-level hard process of $u \bar u$ annihilation  
at the center-of-mass energy equal to  $91.187\,$ GeV of $pp$ collisons at 7 TeV was used.
Note minor imperfections, due to only a discreet set of values 
generated for the  $p_T^Z$ distribution; they are present also in Fig.~\ref{figTest3}.
\label{figTest4}}
\end{figure}

The generated sample has to be convoluted with the initial-state QCD effects. 
For the sake of convolution, each   generated $q \bar q \to Z \to e^+e^- n(\gamma)$ event  is boosted to the laboratory frame
using $x_1$ and $x_2$ of the incoming quarks generated by {\tt WINHAC} Monte Carlo \cite{Placzek:2003zg,Placzek:2009jy}.
To emulate transverse momentum  of $Z$ we use  smearing on the basis of 
a $p_T^Z$ distribution (histogram) 
obtained from  the {\tt RESBOS}  program \cite{Balazs:1997xd}, which was found to model
data within the $4\%$ accuracy, as concluded in Ref.~\cite{phist}.
From this distribution we randomly generate $p_T^Z$ of the $Z$ boson, to be used in the boost 
to the laboratory frame. This is a crude  approximation which, again, is acceptable 
in evaluation of the considered effect, only because it is expected to be small.
In Fig.~\ref{figTest4},  distributions of $p_T^Z$ 
and $y_Z$ (for $u\bar u$ annihilation) are reproduced for the reference.

Two versions of the results will be compared: the one when the exclusive exponentiation, as embedded in {\tt  KKMC},
is featuring the second-order matrix element (option {\tt CEEX2}) and the another one when it is limited to the first-order only (option {\tt CEEX1}), as defined in Ref.~\cite{Jadach:2000ir}.
The technical advantage of correlated samples will help to control differences for the two options in less populated 
regions of the phase space. The {\tt CEEX1} variant will be implemented with an appropriate weight.  
This is advantageous, as we expect the difference of {\tt CEEX2} and {\tt CEEX1} to be rather small, but at the same time 
limitations due to restricted treatment of the initial-state hadronic interactions are present. We have to limit ourselves
(for the particular run) to a fixed flavour of quarks entering the hard process, moreover the virtuality of the intermediate 
state has to be fixed as well. For the presented plots the incoming $u$-quarks were chosen and the intermediate-state virtuality  
was fixed in {\tt KKMC} generation  to $Z$ boson mass. The results for the incoming 
$d$-quarks essentially coincide with the presented ones.

\section{Definition of observable $\phist$}

As proposed and studied in Refs.~\cite{Vesterinen:2009,Banfi:2011,Banfi:2011dx}, an alternative observable to
study the $Z$ transverse momentum ($\zpt$) is $\phist$, defined~\cite{Banfi:2011} as:
\begin{equation}
\phist \equiv \tan \frac{\phi_{\mathrm{acop}}}{2} \,  \sin\theta^{*}_{\eta} \; ,
\label{eq:phist}
\end{equation}
\noindent 
where $\phi_{\mathrm{acop}} \equiv \pi - \Delta\phi$, $\Delta\phi$ is the azimuthal opening angle between 
the two leptons, and the angle $\theta^{*}_{\eta}$ is a measure of the  scattering angle of the leptons with respect to the proton beam direction in the rest frame of the dilepton system.
The angle $\theta^{*}_{\eta}$ is defined~\cite{Banfi:2011} by $\cos\theta^{*}_{\eta} \equiv \tanh[(\eta^{-} - \eta^{+})/2]$, where $\eta^{-}$ 
and $\eta^{+}$ are the pseudorapidities of the negatively and positively charged lepton, respectively.

The observable $\phist$ is expected to be less sensitive to experimental resolution and it can probe largely the same physics as $\zpt$ for small $\zpt$ or 
$\phist$. 
The theoretical calculations for the $\phist$ are documented in Refs.~\cite{Banfi:2011dm,Banfi:2012du}.
The first experimental measurement of the $\phist$  by the D0 Collaboration~\cite{Abazov:2010mk}.
 demonstrated that the order of magnitude improvements in experimental precision could be achieved with the $\phist$ technique.
The $\phist$ was then employed by the ATLAS and LHCb experiments in recent publications~\cite{phist,lhcb}.
The first measurement of the normalized $\phist$ distribution at $\sqrt{s} = 7$ TeV $pp$ collisions
performed by the ATLAS experiment is very likely one of the most precise measurement at the LHC,
with the total uncertainty at the level of $0.5$--$0.8\%$.
The normalized differential cross section was defined~\cite{phist} as $\diffx$, where $\sigma_{\text{fid}}$ is measured within the fiducial
lepton acceptance defined by the lepton ($\ell$ = $e$, $\mu$) transverse momentum $p_{\rm T}^{\ell} > 20\GeV$, the 
lepton pseudorapidity $|\eta^{\ell}| < 2.4$ and the invariant mass of the lepton pair: $66\GeV<m_{\ell\ell}<116\GeV$.
In ~\cite{phist} three lepton definitions with respect to the QED FSR were used.
A "bare" level defined directly by the final-state leptons after QED FSR will be followed in our paper. 
A "dressed" level is defined if lepton four-momenta are recombined with photons radiated within a cone of 
$\Delta R = \sqrt{(\Delta \eta)^2 + (\Delta \phi)^2} = 0.1$. 
Finally, a "Born" level is defined when four-momenta of the final-state leptons before Monte Carlo simulation of QED FSR are used.
The measured $\phist$ distribution in data after correcting only for all detector effects is at bare level.
To account for QED FSR effects and to obtain the  $\phist$ distribution at Born and dressed level a correction,
computed using {\tt PHOTOS} was applied to the data.

The dominant systematic uncertainty of this measurement is due to $\phist$-dependent modelling of QED FSR in this correction, 
assigned to be 0.3\% by comparing predictions from {\tt PHOTOS} (interfaced in {\tt POWHEG+PYTHIA6}) \cite{Alioli:2010xd,Sjostrand:2006za}  and SHERPA, as studied in Ref.~\cite{oanh}.
This uncertainty is extracted by looking at the ratio between the normalized $\phist$ distribution predictions at the dressed or bare levels and the one at the Born level in
{\tt PHOTOS} and in {\tt SHERPA}.

The uncertainties on this  correction are extracted  by looking at the ratio between the normalized $\phist$
distribution at the dressed or bare levels and the one at the Born level  in {\tt PHOTOS} and in
{\tt SHERPA} (in all cases no detector smearing effects are introduced).

The ratio of the predictions of  {\tt PHOTOS} and {\tt SHERPA} in the case of the electron channel is shown\footnote{Fig. 6.5 of Ref.~\cite{oanh} is available on-line from
 \href{https://cds.cern.ch/record/1503540/}{\underline{this hyperlink}.} }
in Fig. 6.5 of Ref.~\cite{oanh}.  Statistical fluctuations are too large to draw firm conclusions, nevertheless
a hint of  difference (wavy structure) seems present which appear to be more profound 
when comparing the two programs for the bare case. 
One can only partly profit from the statistical correlation of the left and 
right hand side plots of Fig. 6.5 which are constructed from the same sample of events.  

We can investigate  this difference by means of comparing the results from {\tt KKMC}: 
at the first-order mode 
and at the second order mode.
This comparison is expected to 
be sensitive to a pure QED effect. The difference as a function of $\phist$ can be used as an correction 
or as estimation of the 
systematic error for the missing second-order matrix element of QED. 
We present details of the comparison in the following section.

\section{Numerical results}

We combine histograms and plot superimposed distributions for  $\phist$ in the  case when
the first and second-order QED FSR matrix elements are used in {\tt KKMC}.
{\tt KKMC} is not suitable for simulations of events in $pp$ collisions. We overcome this obstacle by
switching off all the QED initial-state radiation in {\tt KKMC},  the final-state radiation is then 
retained only. We replace the incoming electrons with quarks. Generation of $q\bar q \to l^+l^- n\gamma$ 
events at the fixed centre-of-mass energy can be then 
performed. We use  several fixed  quark-pair virtualities close to the $Z$ peak\footnote{
Before collecting numerical results, we have checked that our 
calculation for observable 
$\phist$ is properly defined. We have coded independently the $\phist$ angle and
the event selection criteria in the {\tt FORTRAN} and {\tt C++} programming languages.}.

For the {\tt KKMC}-based plots the following cuts were used. These cuts are similar but not identical
with the one used in experimental studies described earlier. We request
$p_T^{l^\pm}> 20$ GeV,  $|\eta_{l^\pm}|<2.4$, in our case  the region  
$1.37< |\eta_{l^\pm}|<1.52$ was rejected. Finally, acoplanarity for leptons was 
requested to be smaller than $3$. 
We have used bare electrons only. We have investigated several options for the cut on lepton pair mass $m_{ee}$ and not only the default one used by the ATLAS experiment:
 $66$ GeV $< m_{ee} < 116$ GeV. 
The cut removes hard-photon configurations and reduces higher-order QED effects, as can be  
 seen in  Fig.~\ref{figTest3}: from 0.2--0.3\% to below 0.1\%. 
The  sample used for construction of these plots was with the
$Z$ virtuality equal to $M_Z$, but we have  checked  the contributions from  
higher virtualities of an intermediate state as well. Then, 
see Fig.~\ref{figTest3}, the  correction was  higher, even if the $m_{ee}$ cut 
was present -- 
it  approached 2\%.

\begin{figure}[htp!]
\begin{tabular}{ccc}
  \includegraphics[width=0.48\columnwidth]{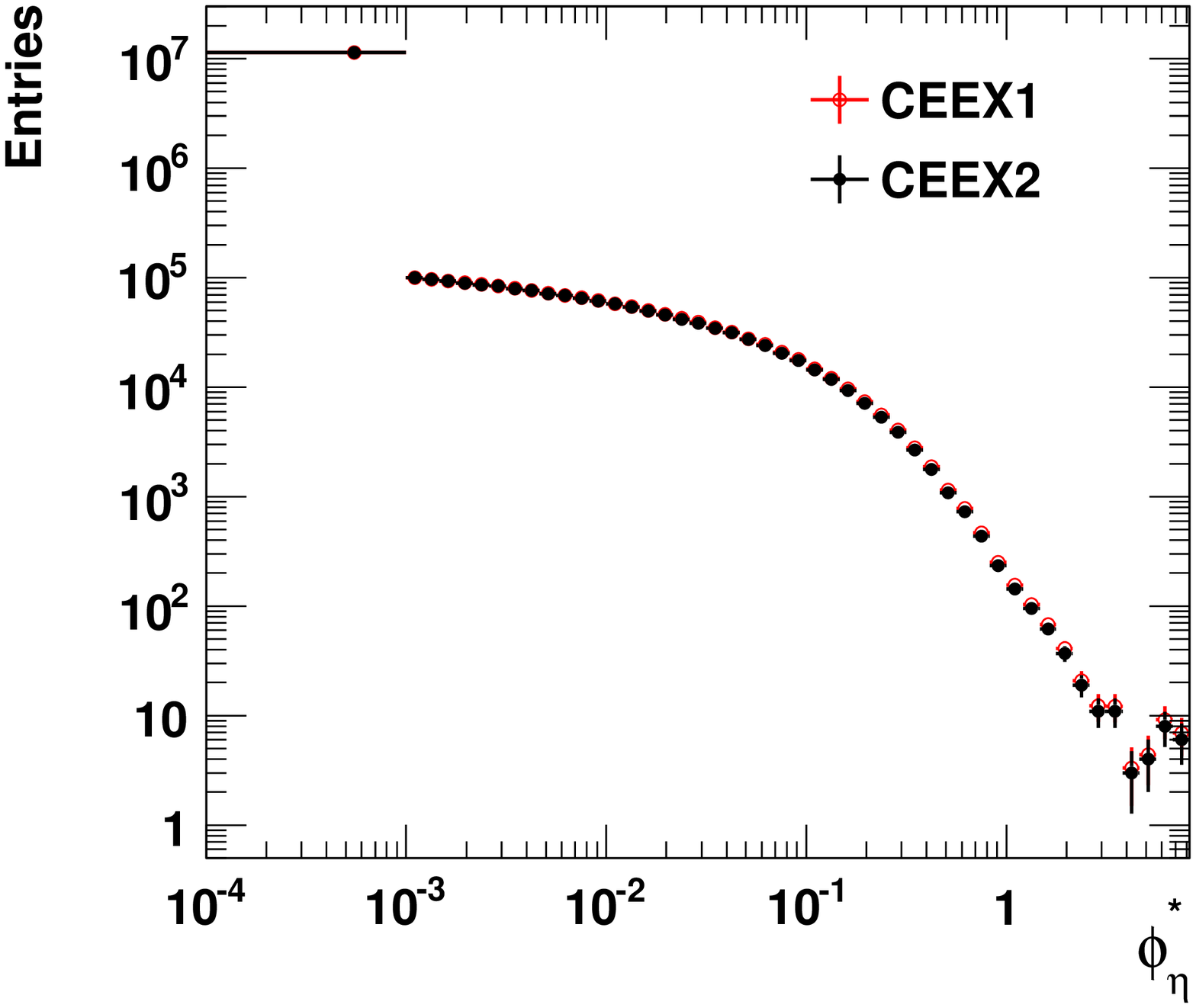} &
  \includegraphics[width=0.48\columnwidth]{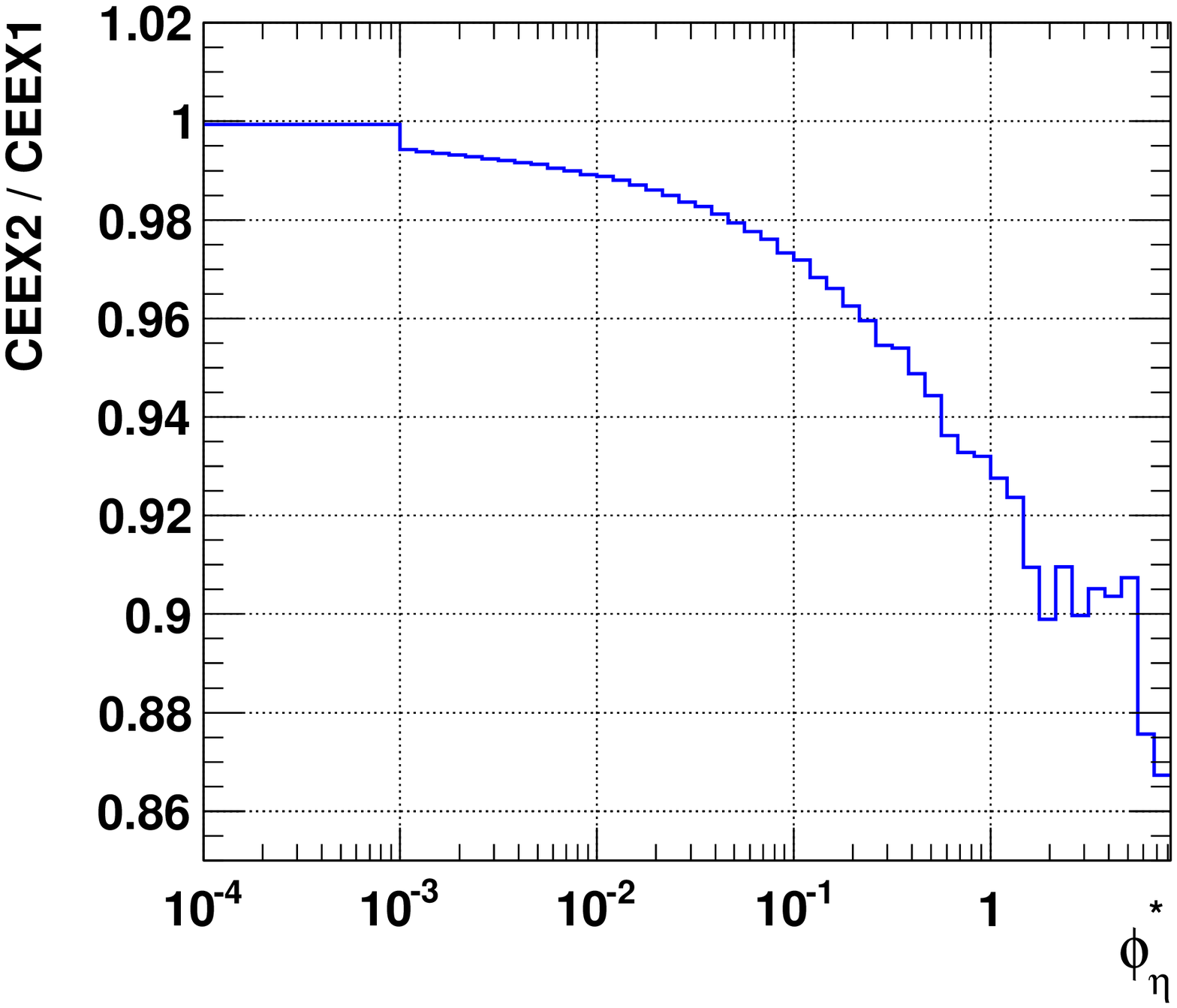}
\end{tabular}
\caption{The $\phist$ distribution: the comparisons of the {\tt CEEX2} and {\tt CEEX1} results.  
The sample of  $u \bar u \to e^+e^- (n\gamma)$ events is generated. The virtuality of the $Z$ boson
equal to its mass
is used.  The boost to the 
laboratory frame of $Z$ is performed prior to histogramming with cuts. 
The longitudinal momentum of $Z$ is generated according to {\tt WINHAC} in $pp$ collision at $\sqrt{s} = 7\,$TeV,
while $p_T^Z=0$.
In the first bin,  the configurations with $\phist$ smaller than $0.001$ are collected. 
\label{figTest1}}
\end{figure}

\begin{figure}[htp!]
\begin{tabular}{ccc}
  \includegraphics[width=0.48\columnwidth]{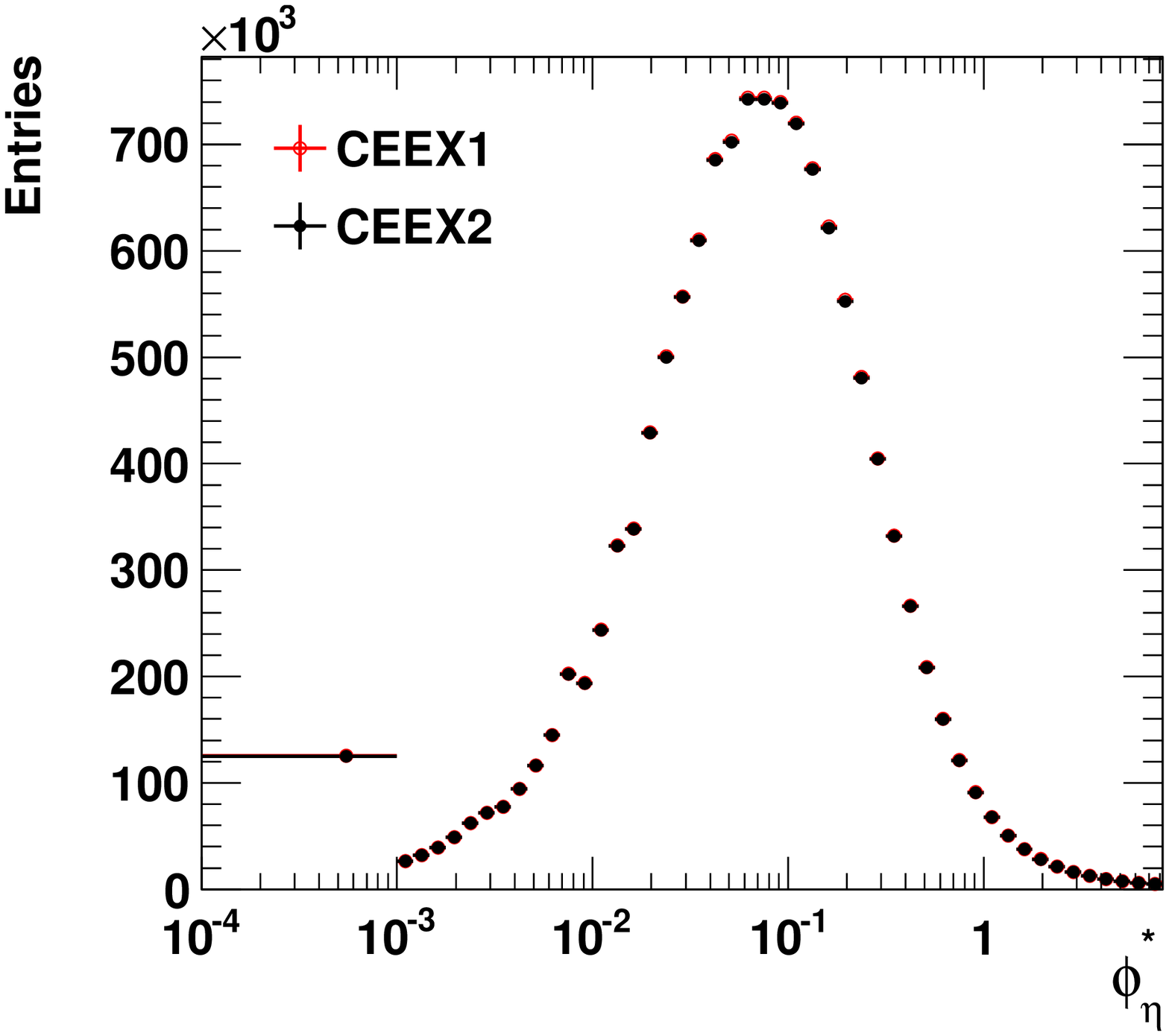} & 
  \includegraphics[width=0.48\columnwidth]{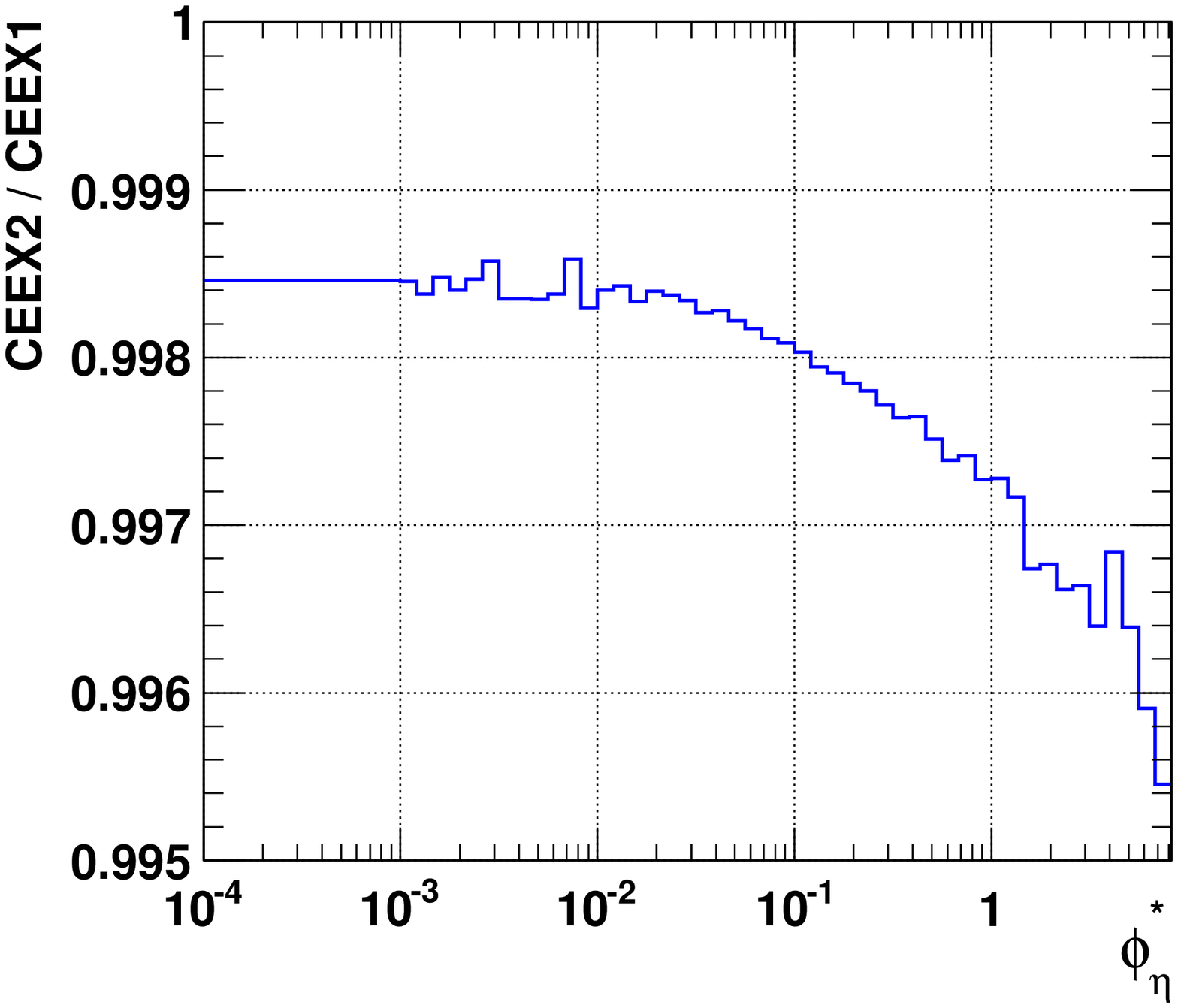} 
  \\ \includegraphics[width=0.48\columnwidth]{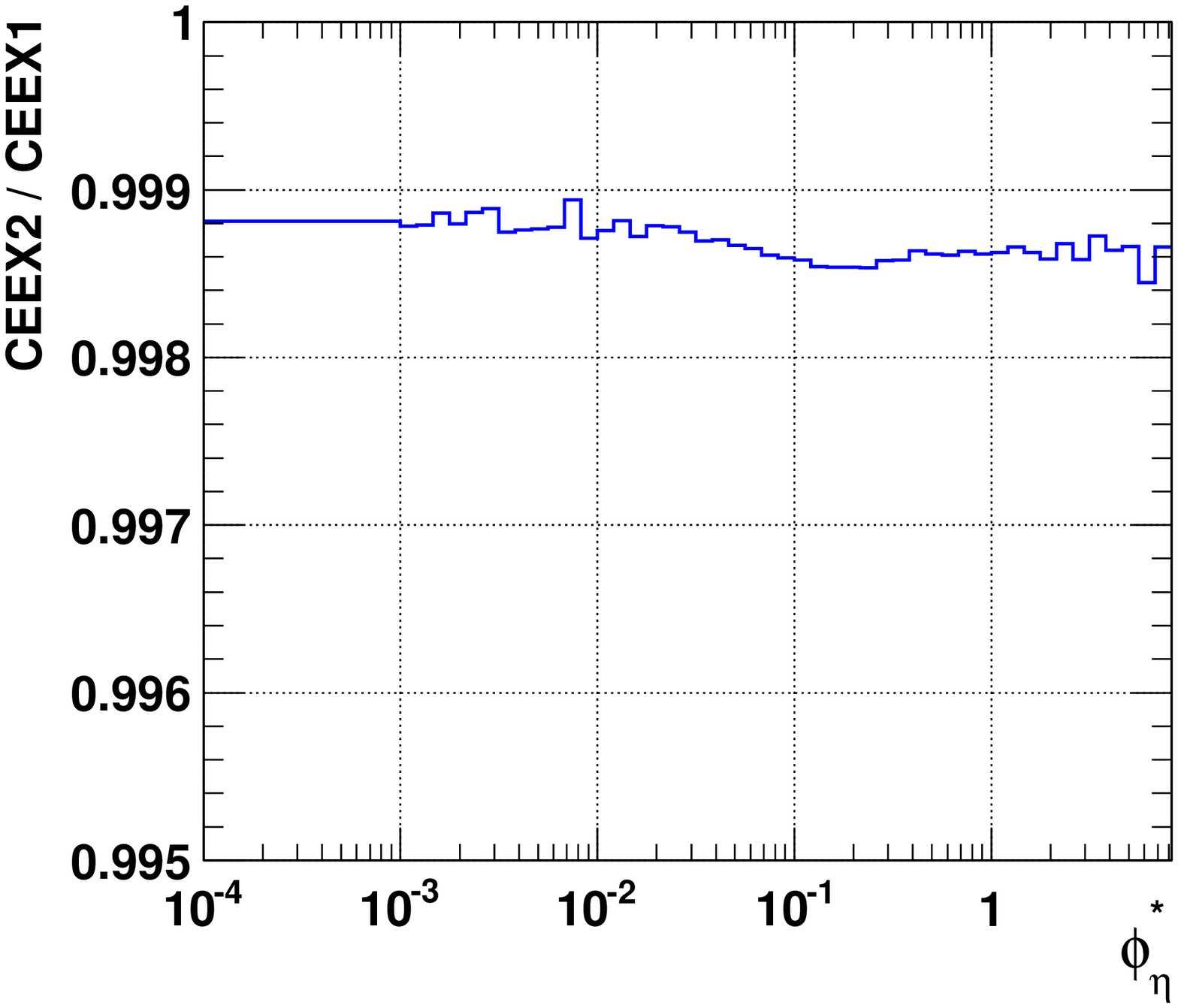} &
  \includegraphics[width=0.48\columnwidth]{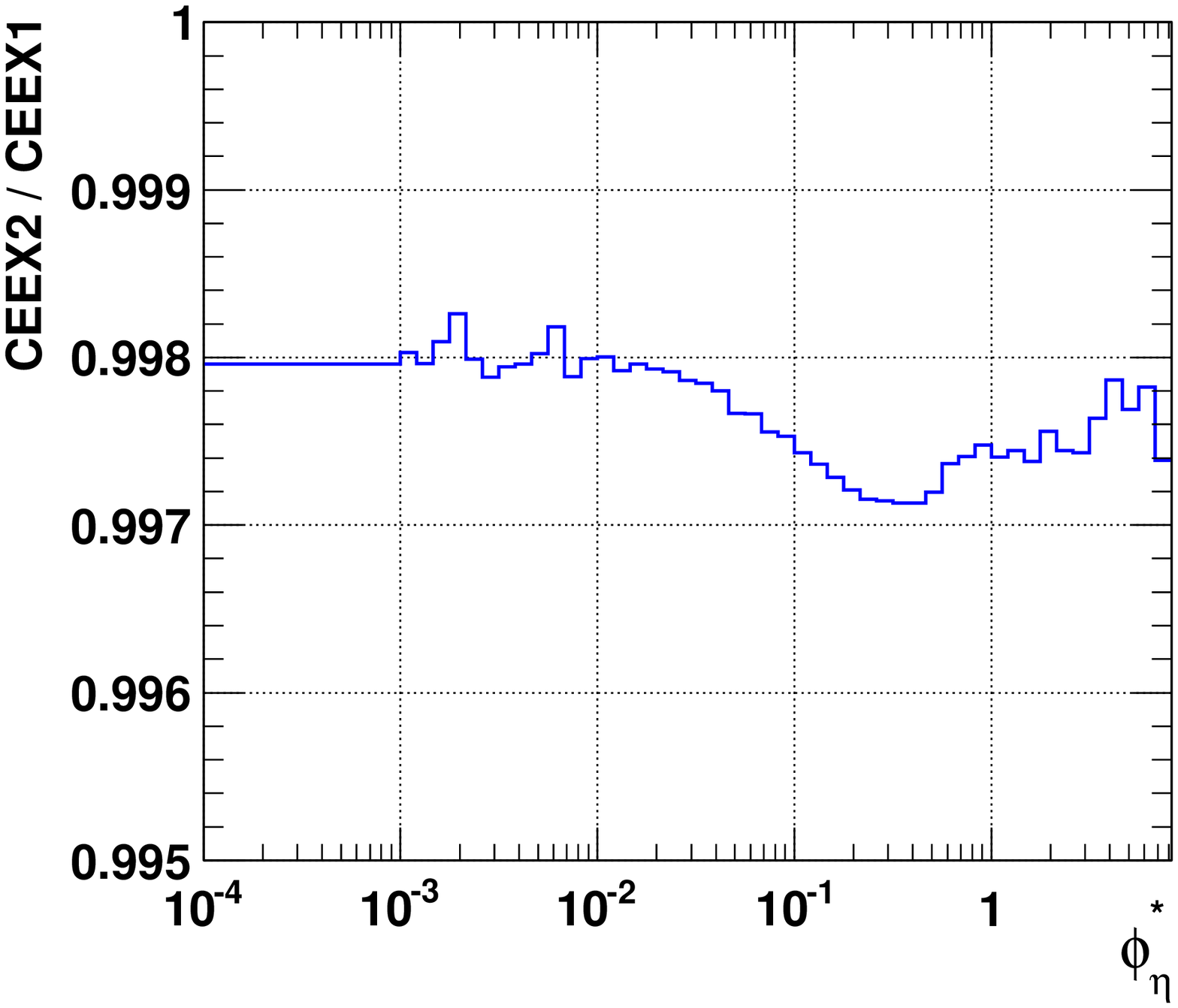} 
\end{tabular}
\caption{The $\phist$ distribution: the comparisons of the {\tt CEEX2} and {\tt CEEX1} results. 
 The sample of  $u \bar u \to e^+e^- (n\gamma)$ events is generated. The virtuality of the $Z$ boson equal to its mass
is used.
The longitudinal momentum of $Z$ is generated according to {\tt WINHAC} in $pp$ collision at $\sqrt{s} = 7\,$TeV, while its
transverse momentum is generated according to the prescription given in the text.
  The boost to the 
laboratory frame of $Z$ is performed prior to histogramming with cuts. 
In the first bin,  the configurations with $\phist$ smaller than $0.001$ are collected. In the upper row of plots, 
cuts as explained in the text except
 cut on $m_{ee}$, are used.
The plot with  $66$~GeV~$< m_{ee} < 116$~GeV is on the bottom-left side.
On the bottom-right side, where in addition to this cut, exceptionally the events with the $Z/\gamma^*$ virtuality  of $115\,$GeV are taken, 
a wavy structure of the correction appears. For the virtuality above $116\,$GeV 
the correction becomes larger,  $2\%$ or more, because the no-bremsstrahlung events do not contribute. 
This demonstrates the limitations of our  
approach. Contributions from all   $Z/\gamma^*$ virtualities have to be combined for realistic prediction, 
rather than of fixed ones, we have used  for the  estimate of the effect size.
\label{figTest3}}
\end{figure}

In Fig.~\ref{figTest1} we show the results with realistic cuts for $\phist$ defined in the previous section, 
where the effects of the 
longitudinal momentum of $Z$ due to {\tt PDF}'s are taken into account. The  transverse momentum, $p_T^Z$, of
$Z$ is, however, set to zero. As a consequence, $\phist$
is non-zero only because  of QED FSR. At least one photon is needed for 
events to populate other bins than the first one. That is why the effects of 
the second-order ME
corrections are numerically large. They represent leading correction: 
nearly all events are placed in
the first bin of the histogram, corresponding to $\phist=0$ or below the detector 
resolution, whereas contents of all other bins are at least two order of magnitude 
smaller. 
The second-order matrix element represent the correction which grows from $1\%$ to 
$15\%$ at the end of the spectrum. As expected, the {\tt CEEX1} prediction is larger
for large $\phist$. The $\beta_1$ correction of the YFS exponentiation is negative 
for hard photons \cite{Jadach:2000ir}.

For Fig.~\ref{figTest3} no restrictions on $p_T^Z$ spectrum were introduced
and one can see that the correction is small over the whole range of $\phist$,
nonetheless grows up to $0.2\%$ at the end of the spectrum.  This is consistent
with what we observe in Fig. 6.5 of
Ref.~\cite{oanh}
at the end of the spectrum for bare
leptons. The distributions of the $Z$ transverse momentum $p_T^Z$ and its rapidity $y_Z$
without any cuts, used in our MC simulations, are shown in Fig.~\ref{figTest4}.

It is technically simple to  prepare such kind of plots for the comparisons 
of {\tt PHOTOS} and {\tt KKMC}, but similar results have recently been presented  in Ref.~\cite{Arbuzov:2012dx}. 
Also the comparisons with other
variants of the matrix element implementation in {\tt KKMC}, such as  
{\tt EEX1, EEX2, EEX3}, may be instructive for detailed
study aiming at the precision implementation of the exclusive exponentiation 
algorithms into
$pp$ collisions Monte Carlo generators. This  is, however, out of the scope of 
this work.

 Let us now comment on  the reliability of presented  results.
  In left hand-side plot of Fig. 6.5 from Ref. \cite{oanh},   the corrections for the $\phist$ calculated
  with dressed electrons, are compared  
for the  {\tt SHERPA} and {\tt PHOTOS} cases; they coincide 
 down to about $0.1\%$. 
The distinct treatment of hadronic initial-state interactions in {\tt PYTHIA} and {\tt SHERPA} can not
therefore be the prime source of  the  wavy structure in $\phist$ corrections calculated  for the  
similar plot but with
 bare electrons, 
see right-hand side of  Fig. 6.5. Note that the two plots of Fig. 6.5 are in part statistically correlated as they are 
constructed from the same sample of events.

The difference between dressed and bare cases likely 
originates
 from the different treatment of   collinear photons.
We may deduce from Fig. 6.5 and our Fig. \ref{figTest3}  
that absent in {\tt SHERPA}  second-order QED 
FSR matrix element can  in part be  responsible for the observed hint   of the difference
in  corrections calculated for bare electrons with the two programs. 
In {\tt PHOTOS} second-, and higher-order 
QED FSR terms, 
important in collinear limits, are taken into account thanks to iterative algorithm.

\section{Summary}

In this paper we have studied the effect of the second-order matrix element with respect to 
 the first-order one. 
We have concentrated mainly on the case when  bare  electrons were used,  
because
it was contributing
to the discussion of the sytematic error in measurement of Ref.~\cite{phist}. 
The Monte Carlo simulation based on the exponentiation was used   for the QED final-state radiation.
The $\phist$ distribution  in $Z/\gamma \to e^+e^- n(\gamma)$ production and decay at LHC was studied.
 We have started with the case
when  $p_T$ of the decaying $Z$ was set to zero, see Fig.~\ref{figTest1}, the effect of the introduction 
of the second-order 
matrix element was large, at the level of $1$--$15\%$.  However, the bulk of the
distribution resided in the first bin of the histogram of $\phist<0.001$.

When the initial-state $Z$-boson  $p_T^Z$ is taken into account, then
the $\phist$ distribution is representing convolution of the dominant effect of initial-state quark transverse 
momenta and  the  QED FSR correction. The effect of  $p_T^Z$
of the incoming $Z$ dominates.  The corrections due to the second order matrix element 
are smaller  in the case of dressed leptons. This is consistent if   the effect
is due to consecutive collinear photons. Already the  second-hardest photon requires the {\tt CEEX2} matrix 
element to be present in the exponentiation-based MC simulations.
The ratio of the parameters defining the range of the lepton-pair invariant mass $m_{ll}$  
to the   virtuality of the intermediate $Z/\gamma^*$ used in our generation seems to be decisive 
for the details of the shape of the {\tt CEEX2-CEEX1} differences.

We can conclude that starting from the precision better than $0.3\%$ for the $\phist$ distribution, 
the dominant contribution of the second-order QED FSR matrix element has to be included in  
discussion of systematic errors for 
Monte Carlo simulations used for the LHC experiments.  
 
The above effects are slightly larger than  presently aimed  precision for the   
fermion-pair emission or interference, as discussed in Ref.~\cite{Arbuzov:2012dx} 
for example. The $\phist$  is an example 
of relatively inclusive observable, where at the least, 
dominant parts of the second-order QED FSR effects have to be taken into account, 
even if the exclusive exponentiation is in use.

We could observe that the effect of the second-order matrix element was more 
significant in the case of the $\phist$ observable defined with the bare leptons than 
with the dressed ones, again confirming that the dominant effect is of the 
leading-logarithmic nature. That is why, one may expect that for many applications,
the systematic error concluded in Ref.~\cite{phist} is an overestimation. 
To understand when it is actually the case,  more detailed studies of cut-off dependence 
than presented in this paper  are necessary.

\section*{Acknowledgements}
Part of this work has been inspired by the discussion with Lucia di Ciaccio, Elzbieta Richter-Was, Stanislaw Jadach and  members 
of the ATLAS LAPP-Annecy group;
 continuous encouragement and comments on intermediate steps of the work are  acknowledged. 
The work is  supported in part by 
the Polish National Centre of Science Grants 
No.\ DEC-2011/03/B/ST2/00220 and DEC-2012/04/M/ST2/00240,
and 
by the Programme of the French--Polish Co-operation between IN2P3 and COPIN within the Collaborations 
Nos.\ 10-138 and 11-142.

\providecommand{\href}[2]{#2}
\end{document}